\documentclass[aps,prb,twocolumn,floatfix,showpacs,showkeys,amssymb,amsmath]{revtex4}
\usepackage{graphicx}
\usepackage{amsmath}
\usepackage{amsfonts}
\usepackage{amssymb}
\begin{document}
\topmargin -15mm
\def\ba{\begin{array}}
\def\ea{\end{array}}
\def\be{\begin{equation}\begin{array}{l}}
\def\ee{\end{array}\end{equation}}
\def\bea{\begin{equation}\begin{array}{l}}
\def\eea{\end{array}\end{equation}}
\def\f#1#2{\frac{\displaystyle #1}{\displaystyle #2}}
\def\om{\omega}
\def\omm{\omega^a_b}
\def\we{\wedge}
\def\de{\delta}
\def\De{\Delta}
\def\va{\varepsilon}
\def\omb{\bar{\omega}}
\def\la{\lambda}
\def\vv{\f{V}{\la^d}}
\def\si{\sigma}
\def\t{T_+}
\def\v{v_{cl}}
\def\m{m_{cl}}
\def\n{N_{cl}}
\def\bi{\bibitem}
\def\c{\cite}
\def\sa{\sigma_{\alpha}}
\def\ua{\uparrow}
\def\da{\downarrow}
\def\mua{\mu_{\alpha}}
\def\ga{\gamma_{\alpha}}
\def\g{\gamma}
\def\ora{\overrightarrow}
\def\pa{\partial}
\def\ov{\ora{v}}
\def\al{\alpha}
\def\bt{\beta}
\def\R{R_{\rm eff}}

\def\muu{\f{\mu}{ed}}
\def\E{\f{edE(\tau)}{\om}}
\def\t{\tau}

\title{AC-field-controlled localization-delocalization transition in
 one dimensional disordered system}

\author{Wei Zhang and Sergio E. Ulloa }
\affiliation{Department of Physics and Astronomy, and Nanoscale
and Quantum Phenomena Institute, Ohio University, Athens, OH
45701-2979 }

\date{\today } 

\begin{abstract}
Based on the random dimer model, we study correlated disorder in a
one dimensional system driven by a strong AC field. As the
correlations in a random system may generate extended states and
enhance transport in DC fields, we explore the role that AC fields
have on these properties.  We find that similar to ordered structures,
 AC fields renormalize the
effective hopping constant to a smaller value, and thus help to
localize a state. We find that AC fields control then a
localization-delocalization transition in a given one dimensional
systems with correlated disorder. 
The competition between band renormalization (band collapse/dynamic localization),
Anderson localization, and the structure correlation is shown to result in interesting
transport properties.

\end{abstract}
\maketitle

\section{Introduction}
The dynamics of electrons in ordered semiconductor superlattices driven by
electric fields has received a great deal of attention. For
example, well-known predictions for quantum mechanical behavior,
such as Wannier-stark ladders and Bloch oscillations, which are
difficult to observe in ordinary solids, were verified in
beautiful experiments. \c{exp1} Furthermore, behavior such as
negative differential conductance, \c{con} fractional
Wannier-Stark ladders, \c{fws} excitonic Franz-Keldysh effect,
\c{fke} among others, have been the focus of recent work. Of
particular recent interest are the localization and delocalization
behavior of electrons in the presence of external electric fields,
which have direct effect on the macroscopic transport properties
of the system. For example, in a system with both AC and DC
fields, an appropriate AC field will delocalize the Wannier-Stark
ladder states induced by a strong pure DC field. \c{hol} An
intense AC field can by itself also lead to {\em dynamical
localization} of the carriers. \c{dy}  This phenomenon has been
studied extensively in various systems, \cite{DL_review}
especially in quantum dot pairs, \cite{Wei-JoseM-Sergio} and
finite linear arrays. \cite{JoseM-Sergio-Nelson-PRB}

Certainly, disorder or imperfections are unavoidable in a real
system. It is widely known that localization due to disorder plays
a fundamental role in a variety of physical situations. In
particular, it has been of interest to investigate disordered
systems in the presence of electric fields: Hone {\em et al}.,
\c{hone2} and Zhang {\em et al}. \c{zzz} studied the case of one
impurity in the presence of AC fields. Holthaus {\em et al}.
studied AC-field--controlled Anderson localization in disordered
semiconductor superlattices. \c{hol1}  As scaling theory shows that all
eigenstates in disordered one-dimensional (1D) systems are
localized, \c{and} previous studies have focused on the effects of
electric fields on the localization length. \c{hol1}

On a related area, the existence of metallic states in a class of
conducting polymers, such as polyaniline and heavily doped
polyacetylene, was identified by  Dunlap {\em et al}. \c{dunlap} 
with {\em extended} states in 1D systems that exist if
short-range correlations in the disordered structure are taken
into account.  The existence of extended states  in this {\em random dimer model} was 
also verified in experiments with
GaAs-AlGaAs superlattices designed to exhibit such 
correlated disorder. \c{exp}

More general correlations have also been studied in 1D systems.
Perturbation theories for the random dimer model were developed in
Ref.\ [\onlinecite{flores}], and particle transport in models with
correlated diagonal and off-diagonal disorder were discussed also
by Flores. \c{flo} The random dimer model driven by a DC field was 
studied in \c{dcdimer}.
The delocalization behavior in 1D models with
long-ranged correlated disorder for the on-site energies was
studied by a renormalization technique,\c{ren} and by a
Hamiltonian approach.\c{cor} More recently, the Kronig-Penney
model with correlated disorder was studied,\c{cor1} demonstrating
that a mobility edge may exist for disordered systems with
appropriate long-range correlated disorder. The role of structural
correlations in the sequence disorder in DNA molecules has also
been studied recently, \cite{wei1,roemer} as this represents a
real system that exhibits structural correlations between the
diagonal and off-diagonal elements in a tight-binding
representation.

In this rich context, it is important to study the competition 
between dynamic localization (due to AC field), Anderson localization
(due to disorder) and the correlation in the disorder, and to investigate the role of
external fields on the transition from the localized to the
delocalized state in 1D systems. In this
paper, we concentrate on the short range correlations of the
random dimer model, and study the localization-delocalization
transition driven by external AC electric fields. We find that AC
electric fields induce a transition from extended to localized
states under suitable conditions, and find the transition point
analytically in the high frequency limit. We also show that the
transition for lower frequencies is shifted in field, as a
precursor of DC-field results. Although our results are for a
relatively simple 1D model potential, we expect that they will be
relevant for a variety of dissimilar systems, including
polymers,\c{dunlap} exciton transfer in active media,
\cite{exciton-hopping} semiconductor superlattices,\cite{exp}
quantum dot arrays, \cite{JoseM-Sergio-Nelson-PRB} and even hole
transport in complex molecules \cite{wei1,roemer}, as their dynamics
is described well by effective 1D models.

After introducing our general model in Sec.\ II and presenting an
analysis of the high frequency regime for the single-dimer system
in Sec.\ III, we present numerical results and general
discussions in Sec.\ IV.

\section{Random dimer model in AC electric fields}

\begin{widetext}
\subsection{Model}
We consider a 1D random dimer model driven by an AC electric
field. The appropriate Hamiltonian is then
 \be H=\sum_m \{\va_m a_m^+
a_m+R(a_{m+1}^+a_m +a _m a^+_{m+1})+medE(t) a_m^+a_m \} ,
 \ee
 \end{widetext}
where $R$ is the hopping amplitude between nearest neighbors,
$E(t)=E_1 \cos(\om t)$ is the time-dependent field with frequency
$\om$, $d$ is the constant separation between chain sites, and the
on-site energy parameter is $\va_m=\va_a$ (or $\va_b$) with
probability $Q$ (or $1-Q$; here we typically choose $Q=1/2$, as it
represents the most disordered system, although other values are
also used), and $\va_b$ is assigned to a {\em pair} of nearest
neighbor sites when it occurs. Since the Hamiltonian is periodic in time, the
Floquet theorem implies that the state can be written as
 \be \psi (x,t)=e^{-i\va t} \sum_n C_n(t) \phi_n(x),
 \ee
where $\va $ is the quasi-energy, $\phi_n$ is the Wannier state
and $C_n$ is the probability amplitude for an electron on site $n$
at time $t$, which is periodic in time, i.e., $C_n(t)=C_n(t+T)$,
with $T=2\pi/\om$. The Schr\"odinger equation
 \be i\f{\pa}{\pa t} \psi(x,t)=H \psi(x,t)
 \ee
can then be written as
 \begin{widetext}
 \be i\f{\pa}{\pa t} C_n(t)=(\va_n-\va) C_n(t)+ R(C_{n+1}+C_{n-1})
 +ned E_1 \cos(\om t) C_n \, . \label{4}
 \ee
Since the term containing the electric field is proportional to
$n$, it is not suitable to perform perturbative calculations. We
introduce the following transformation
 \be C'_n(t)=C_n(t) e^{i n\bt \sin(\om t)} \, ,
 \ee
where $\bt=edE_1/ \om$. It is easy to see that
$|C'_n(t)|=|C_n(t)|$, $C'_n(t+T)=C'_n(t)$, and that $C'_n(t)$
satisfies the equation
 \be
i\f{\pa}{\pa t} C'_n(t)=(\va_n-\va)C'_n(t)+R(e^{-i\bt \sin(\om t)}
C'_{n+1}+e^{i\bt \sin(\om t)} C'_{n-1}) .
 \ee
Since $C'_n(t)$ is periodic in time, we can expand it in Fourier
series, $C'_n(t)=\sum_m A_n^m e^{im\om t}$. Using the identity
\cite{REF}
 \be
e^{i z \sin(\theta)}=\sum_{m=-\infty}^{+\infty} (-1)^m J_m(z)
e^{-im\theta} \, ,
 \ee
where $J_m$ is the $m$-th Bessel function, we can obtain an
equation for $A_n^m$
 \be ( \va_n-\va +m \om)A_n^m+R \sum_l
\left [A_{n+1}^{m+l}(-1)^l J_l(-\bt)+A_{n-1}^{m+l} (-1)^l J_l
(\bt) \right] =0 \, .
 \ee

\subsection{High frequency limit}

We analyze here the behavior of our model in the high frequency
regime. Apart from illustrative, this limit is of practical
importance, since much interest exists in systems driven by THz
fields. \c{fke,thz} In the high frequency limit, $A_n^0$ gives the
most important contribution to $C_n(t)$. The equation for $A_n^0$
is
 \be ( \va_n-\va )A_n^0+R\sum_l(A_{n+1}^{l}(-1)^l
J_l(-\bt)+A_{n-1}^{l} (-1)^l J_l (\bt))=0 \, .
 \ee
We first keep only the terms with $l=0$, and obtain,
 \be
(\va_n-\va) A_n^0=RJ_0(\bt)(A_{n+1}^0+A_{n-1}^0) \, . \label{10}
 \ee
This equation indicates simply that in the high frequency limit,
the effect of the AC field is to suppress hopping and change $R$ to an effective
hopping constant $\R=RJ_0(\bt)$.  This is a well-known result in
driven systems. \cite{DL_review}

Let us consider the corrections coming from terms containing
$A_n^{\pm1}$. The relevant equations are
 \be
(\va_n-\va)A_n^0+RJ_0(\bt)(A_{n+1}^0+A_{n-1}^0)+RJ_1(\bt)(A^1_{n+1}
-A^{-1}_{n+1}-A^1_{n-1}+A^{-1}_{n-1})=0 \nonumber \ee \be
(\va_n-\va+\om)A_n^1+RJ_0(\bt)(A_{n+1}^1+A_{n-1}^1)-RJ_1(\bt)(A^0_{n+1}
-A^0_{n-1})=0 \nonumber
 \ee and \be
(\va_n-\va-\om)A_n^{-1}+RJ_0(\bt)(A_{n+1}^{-1}+A_{n-1}^{-1})
+RJ_1(\bt)(A^0_{n+1}-A^0_{n-1})=0 \, . \label{these}
 \ee
These equations (\ref{these}) can be further simplified, and in
conjunction with (\ref{10}), we find that
 \bea
(B_{n+1}-B_{n-1}) \left(1- \left(\f{RJ_0}{\om}\right)^2 \right)=
\f{2R^2J_0J_1}{\om^2}(A^0_{n+1}+A^0_{n-1})\\
+ \left(\f{RJ_0}{\om}\right)^2
(B_{n+3}-B_{n-3})-\f{2R^2J_0J_1}{\om^2}(A^0_{n+3}+A^0_{n-3}) \, ,
 \eea
 \end{widetext}
where we have defined $B_n=A_n^1-A_n^{-1}$. Thus the higher
Fourier component corrections to the effective bandwidth
$\R=RJ_0(\beta)$ are of higher order in $R/\om$, which of course
makes them small in the high frequency limit, $R/\om \ll 1$.  
In this limit, the model with AC field behaves
essentially  as a system {\em without} electric field,
except for a rescaling of the bandwidth given to the lowest order by
$\R=RJ_0\left(\f{edE_1}{\om}\right)$.

In the random dimer model without AC field, the
localization-delocalization transition occurs when
$\va_-=|\va_a-\va_b|$ is twice the bandwidth. \c{dunlap} We then intuitively
expect that $\va_-=2\R$ would be the transition point between
localized and delocalized states in the presence of the AC field.
Before giving numerical evidence for this transition in the full
random dimer system, we analyze the single impurity case to gain
further understanding of this problem.

\section{Single impurity-dimer case}

It is instructive to see what happens when only one impurity-dimer
is involved in an otherwise periodic 1D chain.  We consider both
the cases with and without an AC field for comparison.

\subsection{Static case}

Let us first consider the scattering effects introduced
by a single {\em site}-impurity in a chain, in the
absence of AC field. \c{dunlap} We let all site energies be
$\va_a$, except for site 0 where it is $\va_b$. From the
eigenvalue equation
 \be (\va_n-\va)
 C_n+R(C_{n+1}+C_{n-1})=0,
 \ee
we can get the transmission probability
 \be
 |T|^2=\f{(4R\sin kd)^2}{\va_-^2+(4R\sin kd)^2} \, ,
 \ee
where $k$ is the wave vector of the incoming Bloch wave, and
$\va_-=|\va_b-\va_a|$ measures the impurity {\em detuning}, and as
such is a measure of the ``disorder" or impurity strength. One can see that for this
one-impurity case, $|T|^2<1$ for $\va_- \not=0$. In a random
multi-impurity system, a series of $n$ scattering events would
naturally lead to $|T|^{2n} \ll 1$, and very small amplitude for
the outgoing wave, resulting in localization in the thermodynamic
limit.

If instead, we assign a pair of sites 0 and 1 with energy $\va_b$
to form a single dimer impurity,
the transmission probability is \c{dunlap}
 \be
|T|^2=1-\f{\va_-^2(\va_-+2R\cos kd)^2}{ \va_-^2(\va_-+2R \cos
kd)^2 +4R^2 \sin^2kd} \, .
 \ee
In this {\em one impurity-dimer} case, $|T|=1$, whenever $\va_-=-2 R
\cos kd$. This is a sort of resonance effect due to the internal
structure of the impurities. Thus, in the presence of the peculiar
kind of short-range correlated disorder described by the random
dimer model, there are states with unity transmission probability,
which clearly have an extended character (even if they only appear
at a single value of the energy). This one impurity-dimer
calculation, makes intuitive  the appearance of extended states
in the random multi-dimer system at peculiar energy values, 
and captures qualitatively the reason for the
unusual behavior of the random dimer model 
[\onlinecite{dunlap,rad1}].

\subsection{The case with AC electric field}

We now turn to explore what happens when an AC electric field is
turned on. Hone {\em et al}. studied the system of an isolated
defect driven by strong electric fields. \c{hone2} We will make
use of Green's functions in terms the Floquet formalism. We
consider the resolvent operator as a function of the complex
frequency $z$
 \be G(z)=\f{1}{z-H}=G^0(z)+G^0(z)VG(z) ,
 \ee
where $H=H_0+V$, $H_0$ is the unperturbed Hamiltonian, $V$ is the
impurity potential, and $G^0(z)=\f{1}{z-H_0}$. In the
representation of Wannier states, the Green's function is
 \be
G_{jl}=G^0_{jl}+\sum G^0_{jj'}V_{j'l'}G_{l'l} \, .
 \ee
In the high frequency limit, $R/\om \ll 1$, $G^0$ is \c{hone2}
 \be
G^0_{jl}=\f{1}{\R} \f{q^{|l-j|+1}}{1-q^2} \, ,
 \ee
where $\R=RJ_0\left(\f{edE_1}{\om}\right)$, and
 \be
q=\f{z}{2\R}\pm \sqrt{\f{z^2}{4\R^2}-1} \, .
 \ee
The sign is chosen such that $q$ falls inside the unit circle. For
an isolated defect with $V_{jl}=\nu \de_{j,0} \de_{0,l}$ (where
$\nu=\va_-=|\va_b - \va_a|$ is the case of different site energy),
the probability $p(l)$ for the defect state (with energy $\va=\R
(q+\f{1}{q})$) to occupy the $l^{\rm th}$ site is determined by
the residue of $G_{ll}$. One finds that \c{hone2}
 \be p(l)=\f{\nu}{\sqrt{\nu^2+4\R^2}}\left(\sqrt{\f{\nu^2}{4\R^2}+1}-
|\f{\nu}{2\R}|\right)^{2|l|} \, .
 \ee  
Notice that $p(l)$ falls exponentially from the site $l=0$ where
the defect is localized, with a characteristic decay length that
is reduced for increasing $|\nu|/\R$, as one might suspect.

Now we consider a one-dimer model, with $V_{jl}=\nu
(\de_{j0}\de_{0l} +\de_{j1}\de_{1l})$, and $\nu=\va_-$. After some
calculation, we find
 \bea
G_{ll}=G_{ll}^0+\nu G_{l0}^0G_{0l}+\nu G_{l1}^0 G_{1l}\\
G_{0l}=\f{\nu G_{10}^0G_{1l}^0+(1-\nu G_{00}^0)G_{0l}^0}{(1-\nu G_{00}^0)(1-\nu G_{11}^0)
-\nu^2 G_{10}^0G_{01}^0}\\
G_{1l}=\f{\nu G_{10}^0G_{0l}^0+(1-\nu G_{00}^0)G_{1l}^0}{(1-\nu
G_{00}^0)(1-\nu G_{11}^0) -\nu^2 G_{10}^0G_{01}^0} \, ,
 \eea
so that
 \be
G_{ll}=aq+\f{1}{2}\nu a^2 q^{2l}
 \left[\f{(1+q)^2}{1-aq(1+q)}-\f{(1-q)^2}{1-aq(1-q)} \right] \, ,
 \ee
where $a$ is defined as $\f{\nu}{\R(1-q^2)}$. Defining
$\g=\f{\nu}{2\R}$, for the pole at $q_1=\f{1}{2\g+1}$, we get
 \be
p(l)\sim (q_1)^{2l} \, ,
 \ee
and since $q_1<1$, this corresponds to a localized state. For the
pole $q_2=\f{1}{2\g-1}$, one gets
 \be p(l)\sim (q_2)^{2l} \, .
 \ee
When $\nu=\va_- > 2\R$, $q_2 < 1$, and this corresponds to a
localized state with localization length $\sim 1/\ln(2\g-1)$. When
$\g$ approaches 1 from above, the localization length diverges, 
indicating a transition to a delocalized state. Thus the single dimer impurity
in an AC field yields
 the same conclusion as in the high frequency case, and the
transition from localized to extended states occurs at the point
$\nu=2\R$.

\section{Numerical results and discussion}

Most of our numerical calculations were performed on a chain with
1501 sites. We solved equation (\ref{4}) with initial condition
$C_n(t=0)=\de_{n,0}$, and analyze the subsequent development. The
site energies were chosen from a bi-valued distribution,
$\va_n=\va_a$ and $\va_n=\va_b$, with probability $\frac{1}{2}$.
As the site prob remains, the ``degree of disorder" is controlled by 
larger values of $\va_-=|\va_b-\va_a|$, as we will see in what follows.

\subsection{High frequency regime $\f{R}{\om} \ll 1$}

\begin{figure}[tbh]
\includegraphics*[width=1.0\linewidth]{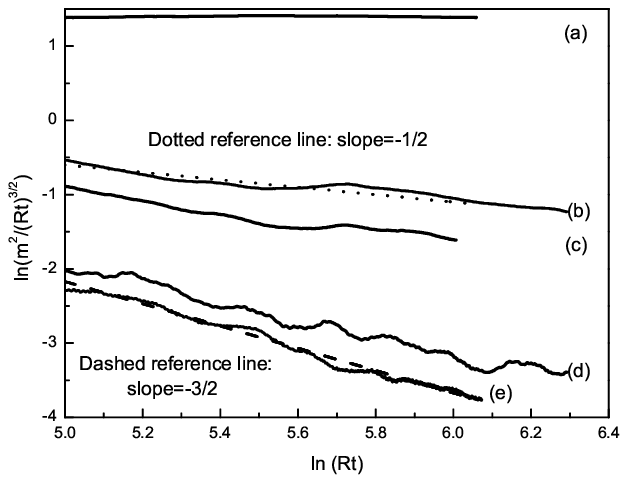}
\caption{The mean-square displacement divided by $(Rt)^{3/2}$ for
varying amounts of disorder in the high frequency regime
$R/\om=0.3$.  Amplitude of AC electric field is
$\beta=edE_1/\om=0.7$; effective hopping constant 
$\R=RJ_0(\beta)=0.8812R$. (a) With $\va_-=|\va_b-\va_a|=\R $
results in superdiffusive transport, $\langle m^2 \rangle \sim
t^{3/2}$; while (b) $\va_-=0.97 \cdot 2\R $, and (c) $\va_-=2\R $
result in diffusive behavior, $\langle m^2 \rangle \sim t$, compared with dotted 
line of slope -1/2.  For
(d) $\va_-=2R $, we see $\langle m^2 \rangle$ is nearly bounded, and for
(e) $\va_-=\om=3.78\R$, 
we find completely bounded $\langle m^2 \rangle$, compared with dashed slop -3/2.}
\label{fig1}
\end{figure}

In Fig.\ 1 we show numerical calculations of the mean-square
displacement, $ \langle {m^2} \rangle=\sum_n n^2 |C_n|^2$, versus
time. One can see in curve (a) that when $\va_-=|\va_b-\va_a|=\R
$, the mean-square displacement $ \langle m^2 \rangle \sim
t^{3/2}$. This is known as the {\em superdiffusive transport}
regime. Diffusive transport ($\langle m^2 \rangle \sim t$) is
shown either when $\va_-=0.97 \cdot 2\R$ (curve b) or perhaps for
$\va_-=2\R$ (curve c).  On the other hand, curve (d) for $\va_-=2R=2.27\R$ shows how the
mean square displacement is increasingly bounded (subdiffusive), $\langle m^2
\rangle \sim t^{0.36}$.  Further increasing $\va_- > 2\R$, as shown
in curve (e)($\va_-=\om=3.78\R$) results
in completely bounded motion $\langle m^2 \rangle \approx$const., as anticipated from the 
analytical discussion above. We believe the subdiffusive behavior for
$\va_-=2R>2\R$ is a crossover behavior due to finite size effects,
masking the anticipated extended $\rightarrow$ localization transition at $\va_-=2\R$.
In fact, from the Eq. (15) (with $R$ replaced by $\R$) for the 
transmission probability, we may estimate the localization length
$\la\sim-1/\ln|T(k)|^2\sim \f{1}{\de+\beta k^2}$, where 
$\va_-=-\de-2\R$, and $\de$ is small. The number of ``extended states"
(states with localization length $\la$ larger than the system size $L$) is
$\De N\sim L(\f{1}{L}-\de)^{1/2}$. We believe theses ``extended states"
lead to the subdiffusive behavior. With increasing of $\de$, as in curve (e),
or increasing system size, one gets $\De N=0$, if $\f{1}{N}<\de$. It is very difficult
to go beyond the crossover regime by numerical simulations, as it requires simulations in very
large system sizes, with longer equilibration times,  requiring  
longer simulation time to obtain accurate values of the self-averaged quantities in a 
$\ln Rt$ fashion. Scaling studies of this transition would be interesting.

The structure of Fig.\ 1 is similar to that shown in
[\onlinecite{dunlap}] in the absence of AC fields. For a fixed AC
electric field amplitude, there is a transition from extended to
localized state behavior with increasing disorder.
The role of AC electric fields can be seen to effectively decrease
the hopping constant, thus contributing to the localization of
carriers. For example, for $\va_-\leq 2R$, in the case {\em without}
electric fields, results in extended states (and diffusive
transport). \c{dunlap} However, when an AC electric field (with
$edE_1/\om=0.7$) is turned on, the mean squared displacement is
suppressed. This localization-delocalization transition is clearly
induced by the AC electric field, as the transition shifts to $\va \simeq 2\R$


It is interesting to see the situation for a stronger field
$\beta=edE_1/\om=2.405$, (the first root of $J_0$). In this case,
$\R=0$, and we expect that even for very weak disorder the state
will be localized. Our results in Fig.\ 2 (with very small
disorder, $\va_-/R=0.33$) show that this is indeed the case. This
is also in agreement with the fact that even in the limit of
$\va_-=0$, i.e. when there is no disorder, the states are
localized when band collapse occurs (i.e. $\R=0$). \c{col} This is
nothing but the well-known dynamical localization. \c{dy} In Fig.\
2 one notices that there are oscillations in the mean square
displacement. This is the manifestation  of the time dependence of
the electric field in this case of weak disorder. In fact these
oscillations also exist in Figs.\ 1 and 3, except that they are
nearly invisible in those cases because the disorder $\va_-$ and
the displacements being larger, ''hide"  the oscillations.

\begin{figure}[tbh]
\includegraphics*[width=1.1\linewidth]{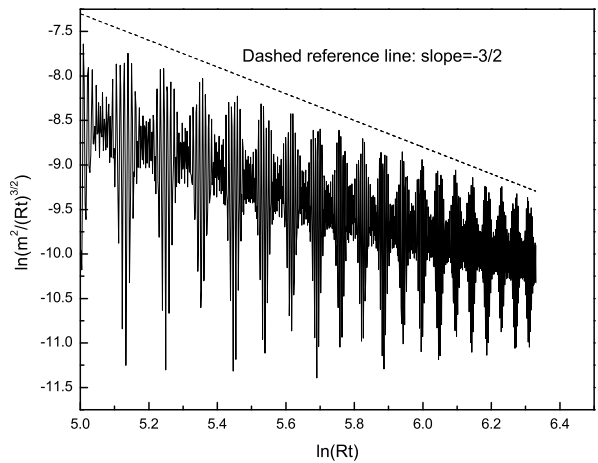}
\caption{The mean-square displacement divided by $(Rt)^{3/2}$ for
$\beta=edE_1/\om=2.405$, $J_0(\beta)=0$, in the high frequency
regime, $R/\om=0.3$. The effective hopping constant is $\R=0$, and
$\va_-=|\va_b-\va_a|=R/3 $.  Notice even weak disorder results in bounded
displacement, characteristic of localization.}  
 \label{fig2}
\end{figure}

\subsection{Low frequency regime $\f{R}{\om}\sim 1$}

\begin{figure}[tbh]
\includegraphics*[width=1.0\linewidth]{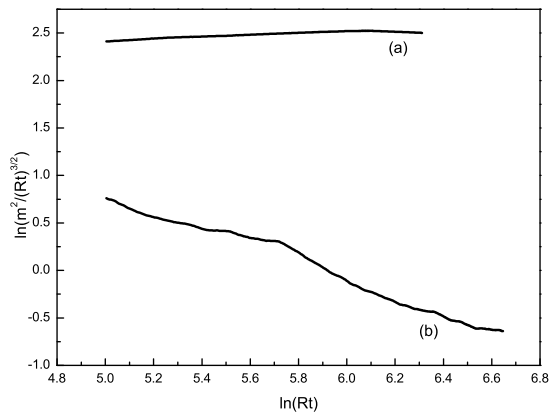}
\caption{The mean-square displacement divided by $(Rt)^{3/2}$ for
varying amounts of disorder in the low frequency regime $R/\om=1$.
The amplitude of AC electric field is $edE_1/\om=0.7$. The
effective hopping constant is $\R=0.8812R$. (a) $\va_-=0.34\R $
shows extended, superdiffusive behavior; (b) $\va_-=\R $ shows near
localization well below the critical value of $\va_-=2\R$ for high
frequency.} \label{fig3}
\end{figure}

In Fig.\ 3 we show the transition from localized to extended state
in the low frequency limit. One can see that the transition point
is no longer the same as in the high frequency regime
$\va_-=2RJ_0(edE_1/\om)$. For example, for $\va_-=\R<2\R$ (curve
b) the state is sub-diffusive, with $\langle m^2 \rangle \sim
t^{0.67}$. As expected, when $\va_- $ is small enough, for example
$\va_-=0.34\R$ (curve a), the state is extended and shows
superdiffusive behavior. These results are of course different
from the high frequency limits, since in this regime our previous
analysis fails. Furthermore, in the extreme low frequency limit,
the system tends to that with a DC field: It is known that
localized states have a power-law behavior in DC field, instead of
the more typical exponential localization in 1D disordered
systems. \c{dc}

\subsection{Inverse participation ratio}

To understand more clearly how the AC electric field controls the
degree of localization,  it is useful to extract information from
the Floquet states for the system driven by periodic electric
fields. The Floquet states $u_m$ can be expanded with respect to
the Wannier states $\phi_l$,
 \be
 u_m(t)=\sum_{l=1}^{N} c_l^{(m)}(t) \phi_l \, .
 \ee
We calculate  the averaged inverse participation ratio $P$,
 \be
P=\f{1}{T}\sum^N_{l=1}\int^T_0 dt |c^{(m)}_l (t)|^4 \, .
 \ee
If a Floquet state is nearly localized at individual Wannier
states, $P$ tends to 1, while $P$ vanishes as $1/N$ if the state
is extended; the larger $P$ characterizes a more localized state.
In Fig.\ 4, we show $P$ for different values of $\va_-=|\va_b -
\va_a|$ and dimer concentrations versus electric field strength
$E_1$ in the high frequency regime, $R/\om=0.1$. 
We find sharp peaks at $edE_1/\om=2.405$, as this value
results in $\R=0$, and thus the effective hopping along the chain
vanishes. We can enhance the degree of localization in the random
dimer model by increasing the detuning $\va_-$ or the dimer
concentration $Q$. For cases (a) and (b) in Fig.\ 4, $Q$ is the
same ($=0.5$), but $\va_-$ changes from 0.16 in (a) to 0.07 in
(b). In contrast, for (b) and (c), the value $\va_-$ is the same,
but $Q=0.2$ is smaller in (c). It is clear that $P$ is larger overall for
the more disordered systems, and although a peak appears always at
$edE_1/\om \simeq 2.4$, decreasing disorder suppresses the peak
value and overall amplitude of $P$. One can also observe that
there is a relatively sudden enhancement of $P$ for the system in
(a) for $edE_1/\om \gtrsim 0.9$, while for (b) and (c) this occurs
between $edE_1/\om \simeq 1.7$, and 3.3. From our previous
discussion, we know that the localization-delocalization
transition occurs at $\va_-=2RJ_0(edE_1/\om)$. From this formula,
we find that the transition point for (a) is in fact at
$edE_1/\om=0.92$, while for (b) and (c) it occurs for
$edE_1/\om=1.78$, and 3.33. These match very well with our
numerical calculation.

\begin{figure}[tbh]
\includegraphics*[width=1.0\linewidth]{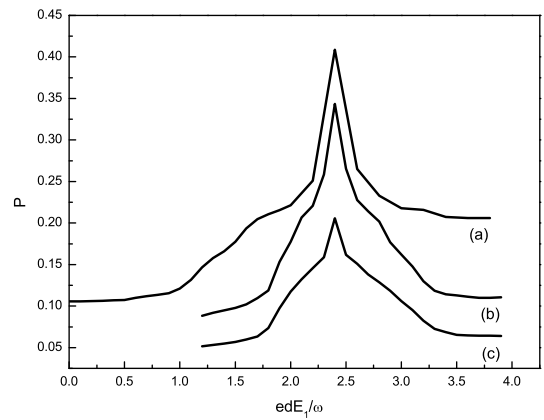}
\caption{The averaged inverse participation ratio versus electric
fields for the random dimer model. Lattice size $N=81$ and
$R/\om=0.1$. (a) $\va_-=0.16  $,  the concentration of dimer of
energy $\va_b$ is $50\%$ ($Q=0.5$); (b) $\va_-=0.07$ and $Q=0.5$;
(c) $\va_-=0.07$, but with concentration of dimer $Q=0.2$. Notice
localized peak at $edE_1/\om = 2.4$ has decreasing amplitude with
decreasing $\va_-$ or $Q$.  Region of high $P$ values agrees with
expected $\va_-=2RJ_0(edE_1/\om)$ (see text).} \label{fig4}
\end{figure}

To elucidate further the role of correlations, we compare $P$ in
an Anderson model (without correlations) with a random dimer model
system, as shown in Fig.\ 5. For a more quantitative comparison,
we let the variance of the Anderson model distribution, $W^2/12$,
be the same as that in the random dimer model,
$\f{1}{2}(\va_a^2+\va_b^2)$. It is evident that $P$ is much larger
in the Anderson model (indicating a more localized system), and
that $P$ varies smoothly with electric field, indicating no
localization-delocalization transition with field. \c{hol1} 
This figure also indicates that an important effect of the presence
of the random dimer short range correlations is to delocalize a
few states, reducing globally the value of $P$ in the system.  It
is clear that the dynamical behavior of a system with correlated
disorder is a subtle competition between correlation and disorder.

\begin{figure}[tbh]
\includegraphics*[width=1.0\linewidth]{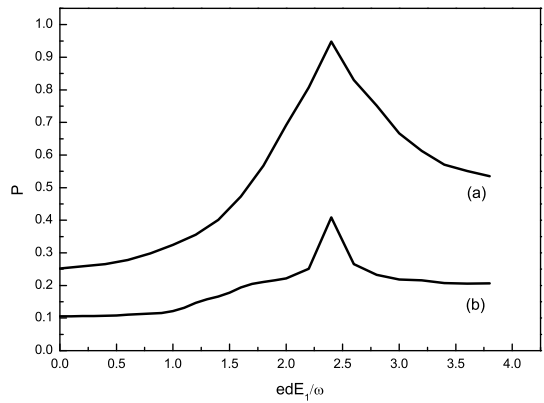}
\caption{The averaged inverse participation ratio versus electric
fields for random dimer model. Lattice size $N=81$ and
$R/\om=0.1$. (a) For Anderson model with
$W=\sqrt{6(\va_a^2+\va_b^2)}=0.392$; (b) for random dimer model
with $\va_b-\va_a=0.16$,  the concentration of dimer of energy
$\va_b$ is $50\%$.  It is clear that the Anderson model with
uncorrelated disorder is more localized and has larger $P$ values
at all fields.}
\end{figure}

\subsection{conclusions}
We have studied the AC-field controlled random
dimer model. The dynamics of our system depends on the competition
between band renormalization (band collapse/dynamic localization),
Anderson localization, and the correlation (dimer structure).
We find that there is an AC electric field
induced transition from extended to localized states, which is
absent in the Anderson model. 
The transition point is found analytically for the high frequency
limit, and found to occur when $\va_-=|\va_b - \va_a|\simeq\R =
2RJ_0(edE_1/\om)$. The dynamical localization is not only recovered as a
natural limit in the absence of disorder, but also shows its effects in 
the transport properties of the system with disorder and correlation (the 
peaks in Figure 4 and 5). The generalization of
our results to a $N$-dimer model is straightforward, and expected
to yield qualitatively similar results. Our theoretical
predictions could be checked in a variety of systems, and
especially on experiments in GaAs-AlGaAs random-dimer
superlattices. \c{exp}  In experiments, tuning external AC field
is a relative easy task compared with changing disorder or correlation
in a desired way. Generalizations to different and more
complex correlations are also expected to give interesting
results.

\begin{acknowledgments}

We acknowledge helpful discussions with J. M. Villas-Boas, and
support from the US-DOE and the 21$^{st}$ Century Indiana Fund.

\end{acknowledgments}

\end{document}